\documentclass[12pt]{article}

\usepackage{epsfig}
\usepackage{amssymb}
\usepackage{amsmath}                          
\usepackage{color}
\usepackage{graphicx}
\usepackage{epstopdf}
\usepackage{setspace}
\usepackage{hyperref}

\def\simge{\mathrel{%
   \rlap{\raise 0.511ex \hbox{$>$}}{\lower 0.511ex \hbox{$\sim$}}}}

\def\simle{\mathrel{
   \rlap{\raise 0.511ex \hbox{$<$}}{\lower 0.511ex \hbox{$\sim$}}}}

\def\s#1{\setbox0=\hbox{$#1$}%
\rlap{\ifdim\wd0>.7em\kern.22\wd0\else\kern.1\wd0\fi /}#1}

\newcommand{\matel}[3]{\langle #1|#2|#3\rangle}
\newcommand{\vev}[1]{\langle #1 \rangle}

\newcommand{\R}[1]{\bar{#1}}
\newcommand{\EH}{E_H}

\begin{document}

\begin{titlepage}
\begin{flushright}\begin{tabular}{l}
Edinburgh/13/08 \\
CP$^3$-Origins-2013-021 \\ 
DIAS-2013-21
\end{tabular}
\end{flushright}

\vskip1.5cm
\begin{center}
  {\Large \bf \boldmath Conformal scaling and the size of $m$-hadrons} 
  \vskip1.3cm 
  {\sc Luigi Del Debbio$^{\,a}$\footnote{luigi.del.debbio@ed.ac.uk}  \&
    Roman Zwicky$^{\,b}$\footnote{roman.zwicky@ed.ac.uk}}
  \vskip0.5cm
  
  $^a$ {\sl School of Physics and Astronomy, University of Edinburgh, 
    Edinburgh EH9 3JZ, Scotland} \\
 
  \vspace*{1.5mm}
\end{center}

\vskip0.6cm

\begin{abstract}
  The scaling laws in an infrared conformal (IR) theory are dictated
  by the critical exponents of relevant operators.  We have
  investigated these scaling laws at leading order in two previous
  papers. In this work we investigate further consequences of the
  scaling laws, trying to identify potential signatures that could be
  studied by lattice simulations.  From the first derivative of the
  form factor we derive the behaviour of the mean charge radius of the
  hadronic states in the theory. We obtain $\vev{r_H^2} \sim
  m^{-2/(1+\gamma^*_m)}$ which is consistent with $\vev{r_H^2}\sim
  1/M_H^{2}$.  The mean charge radius can be used as an alternative
  observable to assess the size of the physical states, and hence
  finite size effects, in numerical simulations. Furthermore, we
  discuss the behaviour of specific field correlators in coordinate
  space for the case of conformal, scale-invariant, and confining
  theories making use of selection rules in scaling dimensions and
  spin.  We compute the scaling corrections to correlations functions
  by linearizing the renormalization group equations.  We find that
  these correction are potentially large close to the edge of the
  conformal window.  As an application we compute the scaling
  correction to the formula $M_H \sim m^{1/(1+\gamma_m^*)}$ directly
  through its associated correlator as well as through the trace
  anomaly. The two computations are shown to be equivalent through 
  a generalisation of the Feynman-Hellmann theorem for the fermion
  mass, and the gauge coupling.
  \end{abstract}

\end{titlepage}

\setcounter{footnote}{0}
\renewcommand{\thefootnote}{\arabic{footnote}}

\tableofcontents

\section{Introduction}
\label{sec:intro}

Gauge theories with an infrared fixed point (IRFP) are studied
currently for building models of strongly interacting electroweak
symmetry
breaking~\cite{Hill:2002ap,Sannino:2009za,Luty:2004ye,Dietrich:2005jn}. At
large distances the couplings flow towards their fixed point values, and the theory
becomes scale-invariant. Theories with an IRFP are said to lie within the {\it
  conformal window}, see e.g. Refs.~\cite{Caswell:1974gg,Banks:1981nn}
for analytical results in the perturbative regime. 

In the absence of supersymmetry, it is difficult to identify a fixed
point in the nonperturbative regime of the theory. Lattice simulations
provide a first principle tool to investigate the low-energy dynamics
of asymptotically free gauge theories. Breaking scale invariance
explicitly, e.g. by introducing a fermion mass term in the action, and
studying the scaling of field correlators as the breaking parameter
tends to zero, has become a common way to characterise IRFPs in
lattice studies e.g. \cite{mCGT-lattice}. A theoretical understanding of the scaling laws is a
necessary tool for these analyses, and a number of useful
(hyperscaling) relations have already been investigated in our
previous work \cite{DelDebbio:2010ze,DelDebbio:2010jy}.  Working out
these scaling relations is an interesting theoretical problem,
independently of its application to the analysis of lattice data. For
a recent discussion of lattice results, we refer the reader to the
comprehensive review that appeared in Ref.~\cite{Giedt:2012it}.

Extending our previous work on mass-deformed conformal gauge theories
(mCGT)~\cite{DelDebbio:2010ze,DelDebbio:2010jy}, we discuss here the
application of the scaling laws to a number of interesting physical
cases, namely the scaling of the hadron size, the scaling corrections,
and the determination of selection rules for field correlators.

The fact that hadrons emerge in a mCGT is a non-trivial empirical
fact.  At least at weak coupling this can be understood as a
consequence of the fermions decoupling below the mass $m$, so that the
low-energy dynamics should be described by a pure Yang-Mills effective
theory, which is believed to be of confining nature
\cite{Miransky:1998dh}.  In practical lattice simulations confinement
is identified through a non-vanishing expectation value for the
Polyakov loop, and it is characteried by the spectrum of the bound
states that determine the correlators of gauge invariant interpolating
fields as in QCD.  In such a theory all hadronic parameters are
controlled to leading order by the coupling $m$, which breaks
explicitly scale invariance, and whose scaling exponent characterises
the long-distance dynamics. This is clearly at odds with the behaviour
observed in QCD, where chiral symmetry breaking requires the Goldstone
bosons to be massless in the chiral limit, while the rest of the
spectrum has a finite mass, which is dictated by some typical hadronic
scale.  We shall refer to the bound states of an mCGT as
\emph{$m$-hadrons} in what follows. For an mCGT the properties of
these $m$-hadrons are very different from the ones commonly
encountered in QCD-like theories. Being able to characterise the size
of $m$-hadrons, and to compute the scaling of the size with the
fermion mass is crucial in order to understand finite-size effects
(FSE) in the results of numerical simulations. When the volume of the
lattice is not large enough to accommodate the $m$-hadrons, FSE
distort the spectrum, and may well obscure the scaling behaviour that
one is trying to identify. This is an important source of systematic
errors in lattice studies, and has been a major concern in the
interpretation of the most recent (and precise) studies of the
spectrum of mCGT, see e.g. Refs.~\cite{Patella:2011kp,Kuti} for a recent
discussions.

There has been a renewed interest recently in the existence of
theories that are scale invariant without being symmetric under the
full conformal group.  For recent work on scale-invariant (SFT) versus
conformal field theories (CFT) see
e.g. Refs.~\cite{Dorigoni:2009ra,Fortin:2011ks,Fortin:2011sz,Fortin:2012hn,Luty:2012ww},
and references therein for earlier work on the subject. It emerges
from our analysis that the scaling laws for field correlators in a
neighbourhood of a fixed point provide a criterion to distinguish SFTs
from CFTs. As a consequence, we discuss the possibility of identifying
the existence of a fixed point describing a CFT by looking at the
scaling behaviour of the correlators when the theory is deformed by a
mass term.

This paper is organised as follows. In Sect.~\ref{sec:confscal} we
rederive briefly the scaling laws, emphasising the features that will
be useful in the rest of our study. In Sect.~\ref{sec:size} we apply
the scaling relations to form factors of conserved currents, and
deduce a scaling law for the radius of the charge distribution inside
the (pseudo)scalar meson. In Sect.~\ref{sec:selection} we use the scaling
laws to formulate a criterion that allows us to distinguish a
scale-invariant theory from a conformal-invariant one as well confining theories. 
Finally in Sect.~\ref{sec:first}, we investigate the subleading corrections to
the scaling laws for generic correlation functions. The corrections are explicitly calculated
for the hadronic mass in two ways and their equivalence is shown using a Feynman-Hellmann type relation for the gauge coupling. 
The relation of the charge radius to the derivative of the form factor is summarised 
in App.~\ref{app:charge} for the readers convenience.

\section{Conformal scaling}
\label{sec:confscal}

Let us concentrate here on a theory with only one relevant
perturbation at the IRFP, whose coupling we denote by $m$, and let us
introduce an UV cut-off  $\Lambda$; $O_1$
and $O_2$ are two local operators. The generic two-point
correlator, evaluated on two arbitrary physical states $\varphi_{a,b}$, in the regulated bare theory:
\begin{equation}
  \label{eq:corr}
  C(x,m,\Lambda) = \matel{\varphi_a} {O_1(x) O_2(0)} {\varphi_b} 
\end{equation}
depends on the distance $x$, the coupling $m$ and the scale $\Lambda$.
In the expression above we rescale the dimensionful coupling $m$ by
some reference scale $m_0$, so that the correlator depends on the
dimensionless coupling $\hat m \equiv m/m_0$.  We denote the
\emph{scaling dimension of the coupling} $m$ by $y_m \equiv d_m
+\gamma_m$ where $d_m$ and $\gamma_m$ are the engineering and
anomalous dimension respectively (and clearly $d_m=1$).  
We shall adopt the same conventions as in
Refs.~\cite{DelDebbio:2010ze,DelDebbio:2010jy}, denoting by $d_{O_i}$ and $\gamma_{O_i}$ the classical and anomalous
dimensions of $O_i$ and therefore the \emph{scaling dimension
  of the operator} $O$ reads: $\Delta_{O_i} \equiv
d_{O_i} + \gamma_{O_i}$.  
For the sake of clarity, anticipating section \ref{sec:first},  we shall denote by $\gamma^*$ (and thus $\Delta^*$) any anomalous dimension at the fixed point in order to distinguish it from the one away from the fixed point.

In computing the leading scaling  we
perform, as usual, an renormalisation group (RG) transformation $\Lambda \to \Lambda/b$:
\begin{equation}
  \label{eq:RG-G}
  C(x,\hat m, \Lambda) = b^{-(\gamma^*_{O_1}+\gamma^*_{O_2})} C(x,
  b^{y^*_m} \hat m , \Lambda/b)  \;, \quad y^*_m = 1 + \gamma^*_m\, .
\end{equation}
followed by a rescaling of  all mass scales by a factor of $b$, 
on the RHS of \eqref{eq:RG-G}:
\begin{equation}
  \label{eq:scale-G}
  C(x, b^{y^*_m} \hat m,\Lambda/b) = b^{-(d_{O_1}+d_{O_2}+ d_{\varphi_a} + d_{\varphi_b}  )} C(x/b,b^{y^*_m} \hat m , \Lambda) \;.
\end{equation}
 A crucial observation is that the physical states are free of anomalous scaling \cite{DelDebbio:2010jy}.
 Combining Eqs.~(\ref{eq:RG-G})
and (\ref{eq:scale-G}) we get
\begin{equation}
  C(x,\hat m, \Lambda)  = b^{-(\Delta^*_{O_1}+\Delta^*_{O_2})}  
  C(x/b, b^{y_m^*} \hat m,\Lambda) \;.
\end{equation}
We can exploit the arbitrariness of $b$ and choose it
such that $b = \sqrt{x^2} m_0$. This then implies that 
\begin{equation}
  \label{eq:Gtof}
  C(x,\hat m,\Lambda) = \left(\hat x^2 \right)^{-\alpha}  \,
  (m_0)^{d_{O_1}+d_{O_2}+ d_{\varphi_a} + d_{\varphi_b} }  \, 
  F(\hat x^{y^*_m} \hat m, \hat \Lambda)
\end{equation}
with $\alpha\equiv(\Delta^*_{O_1} + \Delta^*_{O_2} + d_{\varphi_a} + d_{\varphi_b} )/2$,
$\hat \Lambda = \Lambda/m_0$ and $F$ a dimensionless function. We will use
this particular form of the scaling law to derive some physical
consequences in the following sections. The application to mCGTs
can be inferred indirectly from the caption of Fig.~\ref{fig:3-types}. 
Discussion of finite size effects to Eq.~\eqref{eq:Gtof} can be found in 
appendix \ref{app:genFSE}.

\begin{figure}[h]
  \includegraphics[width=\textwidth]{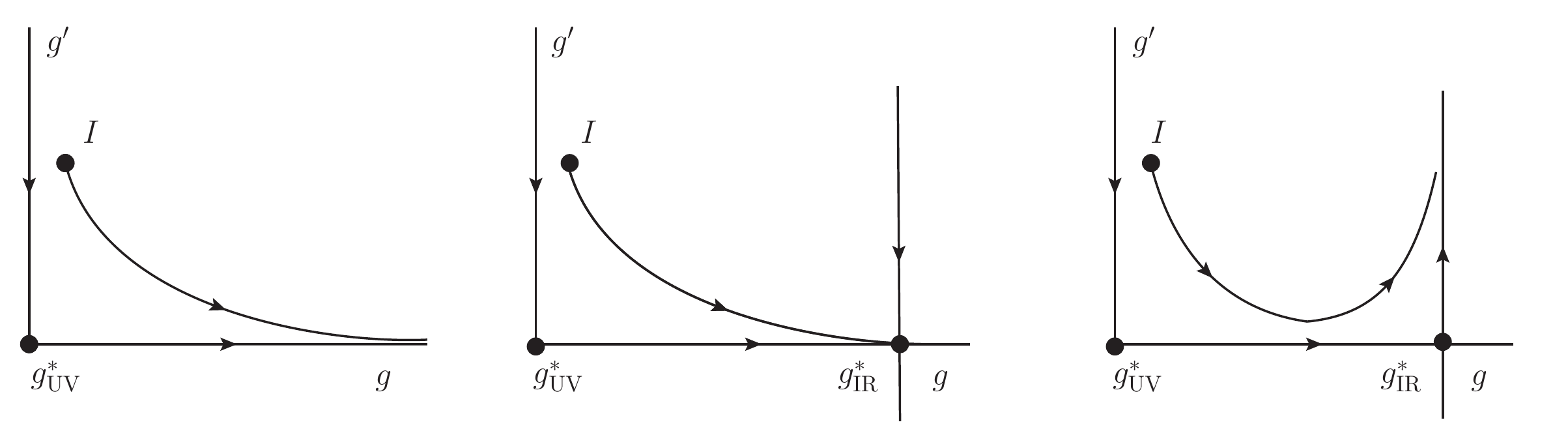}
  \caption{\small Overview of behaviour of relevant and irrelevant
    directions at UV and IR fixed points (FPs).  The couplings, say
    ${\cal L}^{\rm eff} \sim g_i O_i$ with $y_i = 4 - \Delta_i$, 
    fall into relevant ($y_{g_i} > 0$), irrelevant ($y_{g_i} <
    0$) and marginal ($y_{g_i} = 0$) classes at the FPs. In all cases
    there is a trivial $(g^*_{\rm UV} = 0)$ ultraviolet fixed point (UVFP) and the $y$-axis
    corresponds to its critical surface.  (left) The couplings $g'$
    and $g$ are irrelevant and relevant at the UVFP.  An example of
    which is QCD with $g$ being the gauge coupling and $g'$ the quark
    mass.  (middle) non-trivial IR fixed point $(g^*_{\rm IR} \neq
    0)$. The direction $g$ is relevant and irrelevant at UV and IRFP
    respectively whereas $g'$ is irrelevant at both fixed points. An
    example is IR-conformal gauge theory (with $m=0$) and $g'$ is four
    quark operator provided $y_{g'} < 4$ is really the case.  (right)
    The same as before but $g'$ is relevant at IR fixed point. 
    Examples are, assuming $y_m > 0$, mCGT where $g'$ is the mass $m$.
    Another example is IR-conformal gauge theory with $m=0$ where $g'$
    is four quark operator where this time $y_{g'} > 0 $.}
  \label{fig:3-types}
\end{figure}

\subsection{Comment on additional relevant directions in mCGT}
\label{sec:comment}

In derivations like the one shown in the previous section it was
assumed that there is only one relevant operator driving the system
away from the IRFP. Current lattice results seem to suggest that this
is indeed the case for the theories that have been investigated so
far. Nevertheless it might be the case that four quark operators
\begin{equation}
  \label{eq:4fermi}
  {\cal L}^{\rm eff} = \frac{c_{\bar qq \bar qq}}{\Lambda_{\rm ETC}^2} {\bar qq \bar qq} 
\end{equation}
that do appear for example in extended technicolor (TC) models, become relevant,
i.e. $\Delta_{\bar q q \bar q q} < 4$. In this case a situation like
the one shown in Fig.~\ref{fig:3-types} (right)
will apply: in the very far IR this operator will grow and drive the
system away from the fixed point; both the mass of the fermions and
this additional coupling need to be tuned for the system to be on the
critical surface. Academically one could hope to hit a trajectory that
goes directly in the UVFP for which $c_{\bar qq \bar qq}|_{UVFP}=0$,
and then flow out of the UVFP along the renormalized trajectory
flowing into the IRFP. This would be the equivalent of finding a
perfect action for the IRFP. In practice, e.g. when setting the bare
parameters in a simulation at finite lattice spacing, it is impossible
to tune the system exactly to this point. The simple plaquette action
does contain higher dimensional couplings by construction, and an
infinite amount of tuning is needed to find a perfect action. Thus
summa summarum the study of the scaling dimension of higher
dimensional operators within mCGT will remain an important topic in
practice.

\section{Size of $m$-hadrons from form factors}
\label{sec:size}

In this section we characterise the size of hadronic states in mCGTs
by studying the radius of their charge distribution. The radius of the
charge distribution is defined from the derivative of the
 form factor of the state; the latter is defined in
turn from the matrix element of the conserved vector current between
hadronic states. Scaling laws for the derivatives of the form factor
can be deduced from the scaling laws we have obtained for the matrix
elements in our previous paper~\cite{DelDebbio:2010jy}.

In the following let us consider a matrix element where a scalar
particle $H$ probes a conserved vector current.  On the grounds of Lorentz
covariance the matrix element may be parameterized as
follows~\footnote{The current $V_\mu$, which we do not specify any
  further at this point, might be in the flavour singlet or adjoint representation. 
  The main point is that $H$ couples to it. 
  Subtle cases
  in the real hadronic world are $f_+^{\pi_0} =0$ by virtue of
  $C$-covariance; yet $f_+^{K_0} \neq 0$ as $K_0$ is not
  $C$-eigenstate.}
\begin{equation}
  \label{eq:FF}
  \matel{H(p_1)}{V_\mu}{H(p_2)} = (p_1+ p_2)_\mu f^H_+(q^2)   \;, \qquad J^{\rm PC}(H) = 0^{\rm PC}\;,
\end{equation}
where $ q \equiv p_1 - p_2$ is the momentum transfer to the current.
Note that the structure $ (p_1- p_2)_\mu f^H_-(q^2)$ vanishes by
virtue of current conservation: $\partial \cdot V =0$.  The function
$f^H_+(q^2)$ is known as a form factor: its value at zero momentum
corresponds to the charge of $H$ under the current $V_\mu$, and its
derivative corresponds to the square of the charge distribution c.f. appendix \ref{app:charge}.
For instance for the pion form factor in QCD,
\begin{equation}
  \label{eq:pion-observables}
  f^{\pi_\pm}_+(0) = \pm 1\;, \quad 
  \vev{r^2_{\pi_\pm}} =  \left. 6  \frac{d}{dq^2} f^{\pi_\pm}_+(q^2)\right|_{q^2 =0 } \;.
\end{equation}
We wish to emphasise that \eqref{eq:pion-observables} is not related 
to the pion's special role in QCD as should be clear from the notes in the appendix \ref{app:charge}. We shall later on contrast the behaviour of the pion charge radius in  QCD with the charge radius of a generic $m$-hadron.
In order to determine the scaling exponents, following the notation in
\cite{DelDebbio:2010ze,DelDebbio:2010jy}, we define
\begin{eqnarray}
  f_{+,n}^{H} \equiv \left. \frac{d^n}{d(q^2)^n} f^{H}_+(q^2)\right|_{q^2 =0 } \sim m^{\eta_{f_n}} \;,
\end{eqnarray}
and shall assume that the derivatives exist.  Our main interest is to
establish the behaviour of the size of the $m$-hadrons as a function
of the relevant perturbation $m$. We will proceed in two steps: (i) we
derive the relative difference $\eta_{f_{n+1}}-\eta_{f_{n}}$, and
(ii) we determine $\eta_{f_{0}}$.
\begin{itemize}
\item[(i)] The mass dependence of the form factor, $f(q^2) \equiv
  f^H_+(q^2)$ for shorthand, is summarised in a scaling law akin to 
  Eq.~\eqref{eq:Gtof}:
  \begin{equation}
    \label{eq:taylor}
    f(q^2) =
    \tilde{f}(\hat  q^2/ \hat m^{2/y_m^*}) = \tilde{f}(0) +   \tilde{f}'(0)  \left( \frac{\hat
        q^2}{ \hat m^{2/y_m^*}} \right)+ 
    \frac{1}{2} \tilde{f}''(0) \left( \frac{\hat q^2}{ \hat m^{2/y_m^*}} \right)^2 + ... \;,
  \end{equation}
  where the dots stand for higher terms in the Taylor expansion. Note
  there is no dependence on the RG-scale as the current is conserved.
  From Eq.~\eqref{eq:taylor} it is immediate to deduce
  \begin{equation}
  \label{eq:difference}
    \eta_{f_{n+1}} - \eta_{f_{n}} =  -2/y_m^*
  \end{equation}
\item[(ii)] Second we shall show $\eta_{f_{0}} =0$.  It follows
  directly from our master formula \cite{DelDebbio:2010jy}:
  \begin{equation}
    \label{eq:master}
    \matel{\varphi_2}{ O(0)}{\varphi_1}   \sim 
    \left(  \hat m \right) ^{(\Delta^*_O + d_{\varphi_1} + d_{\varphi_2})/y_m^* } 
  \end{equation}
  where $\varphi_{1,2}$ are physical states. We note that $\Delta_{V_\mu} = 3$ (since $V_\mu$ is a conserved current) and that 
  $d_{\varphi_1} = d_{\varphi_2} = -1$ which implies that $f^H_{+,1}(0) (p_1+p_2)_\mu \sim m^{1/y_m^*}$.
  Since the energy momentum vector is free from anomalous scaling 
  it counts like its engineering dimension in the formula in the nominator of the exponent 
  in \eqref{eq:master} and therefore $\tilde{f}(0) = f(0) \sim {\cal O}(1)$ (i.e. $\eta_{f_0} = 0$).
  Another way to arrive at the same result is to notice that $\tilde{f}(0)$ is equal to the charge 
  and since the latter cannot scale with external parameters like the mass 
  this implies that $\tilde{f}(0)$ is independent of the mass and thus $\eta_{f_0} = 0$.
\end{itemize}
Putting the two results together we get:
\begin{equation}
  \label{eq:result}
  \eta_{f_n} = \frac{-2 n}{y_m^*} \equiv  \frac{-2 n}{1+ \gamma^*_m} \;,
\end{equation}
and for the mean charge radius squared \eqref{eq:pion-observables}  we obtain:
\begin{equation}
\label{eq:r2H}
  \vev{r^2_{H}} =  \left. 6  \frac{d}{dq^2} f^{H}_+(q^2)\right|_{q^2
    =0 } \sim m^{\eta_{f_1}} = 
  m^{-2/y_m^*} \sim 
  \frac{1}{M_H^2} \;,
\end{equation}
where $M_H$ denotes the mass of the hadron $H$ and 
we have used the general result $M_H \sim m^{1/y_m^*}$ derived for
the entire hadronic spectrum in Ref.~\cite{DelDebbio:2010jy}. Thus in
summary the size of the $m$-hadrons is inversely proportional
to the hadronic mass.  Whereas this result does not seem surprising
it is of importance for controlling FSE on the
lattice.  Whereas the scaling laws gives information on the relative size
of hadrons  for different values of $m$ it does not determine its absolute size $ \vev{r^2_{H}} = K_{r^2_{H}} M_H^{-2}$. 
The determination of $K_{r^2_{H}} \sim {\cal O}(1)$  could then be
pursued by a  measurement of the slope of the form factor
\eqref{eq:FF} through Eq.~\eqref{eq:r2H}. Using twisted boundary
conditions could help in improving the momentum resolution, and hence
in resolving better the slope of the form factor.  The discussion of finite size effects
in the context of the form factor can be found in appendix \ref{app:chargeFSE}.

It would seem that the arguments of the  form factor of a scalar coupled to a conserved current \eqref{eq:FF} ought to generalise to higher spin hadrons. The application 
to the analogue of the proton electromagnetic form factor should be rather straightforward.
In general a more detailed analysis would necessitate the consideration of 
the corresponding polarisation tensors.  Suppose two higher spin hadrons
couple to an operator $O$ that is not necessarily related to a physical charge.
Even though $\eta_{f_0}(O) \neq 0$ in general, 
we anticipate that  the extension of the overlap with the operator $O$ is determined by \eqref{eq:difference}  based \eqref{eq:taylor} which in turn follows from generic scaling arguments.

Let us briefly open a parenthesis here.
Since  $M_H \approx  K_{M_H}  m^{1/y_m^*} \Lambda_{\rm ETC}^{1-1/y_m^*}$ with   
$m \ll \Lambda_{\rm ETC} $ (c.f. Fig.~\ref{fig:coupling} for an explanation of $\Lambda_{\rm ETC}$), $K_{M_H}=\mathcal{O}(1)$, one concludes that for 
\begin{equation}
  \label{eq:hierarchy}
  y_m^* \equiv 1+ \gamma^*_m > 1  \quad \Rightarrow \quad  m < M_H \;, 
\end{equation}
at least for sufficiently small $m$ to overcome the unknown   ${\cal O}(1)$-coefficient 
discussed above. Other than that the hierarchy is controlled by the positivity of $\gamma_m^*$ which is of course dependent on the actual gauge theory.  Furthermore whereas 
the unitarity bound implies $\gamma_m \leq 2$ no lower bound exists other than the fact 
that for $\gamma_m < -1$ the operator becomes irrelevant which goes against our working 
assumption as well as all results, known to the authors, in the literature.  

It is interesting to contrast the behaviour of the mean charge radius
of the (pseudo)scalar meson in mCGT to the one obtained for the Goldstone boson in QCD.  More precisely, since in both cases the masses vanish in the limit $m \to 0$ it is clear from a heuristic viewpoint that, for a state with sharp momentum, the particle cannot be localised 
and therefore one expects the charge radius to diverge. The functional behaviour of the divergence is though not clear a priori.
In a theory where chiral symmetry is spontaneously broken, the dynamics of
the light Goldstone bosons is described by chiral perturbation
theory. The mean charge radius can be computed in perturbation theory,
and it is found to diverge logarithmically with the pion mass \cite{GL85.III}. This
difference suggests that the existence of a conformal fixed point
could be characterised by studying the scaling of $\vev{r^2_{H}}$ for
the pseudoscalar meson.  We wish to reemphasize \cite{DelDebbio:2010ze} that the scaling laws for the
mass parameter imply that there is no remnant of the pion as a pseudo Goldstone boson  in a mCGT.


\section{Exploiting selection rules of CFT-correlators}
\label{sec:selection}

We discuss in this Section how to exploit selection rules for
two-point vacuum correlators originating from scaling dimensions and
spin of the quasi-primary (to be commented on further below) operators.  In
subsection~\ref{sec:CFTvsSFT} we contrast these aspects from the
viewpoint of distinguishing CFTs from SFTs (c.f. \cite{Nakayama:2013is} for lecture notes on this topic), while in
subsection~\ref{sec:CFTvsConf} we focus on differentiating conformal
from confining behaviour.

\subsection{CFT vs SFT}
\label{sec:CFTvsSFT}

Consider first a scale invariant theory, and specifically
(quasi)-primary fields ${\cal O}_{1,2}$ and ${\cal O}_3^\mu$ with
respective scaling dimension $\Delta^*_{{\cal O}_1} = \Delta^*_{{\cal
    O}_3} \neq \Delta^*_{{\cal O}_2}$. In the absence of symmetry
breaking, the short distance correlator obey the following selection
rules:
\begin{enumerate}
\item \emph{\underline{Scaling dimension}}  \cite{Ginsparg:1988ui}\footnote{For more elaborate
  forms under open flavour and Lorentz indices of the SFT correlators
  we refer the reader to references
  \cite{Fortin:2011sz,Fortin:2011bm}.}
\begin{equation}
  \label{eq:S-CFT-G}
  C(x) = \matel{0}{{\cal O}_1(x) {\cal O}_2(0)}{0}  \sim
  \begin{cases}
    (x^2)^{-\alpha}\phantom{0}  \quad \text{SFT} \\[0.1cm]
    0\phantom{(x^2)^{-\alpha}} \quad   \text{CFT}   \end{cases} \;, 
\end{equation}
with $\alpha \equiv (\Delta^*_{{\cal O}_1}+\Delta^*_{{\cal O}_2})/2$.  
\item \emph{\underline{Spin}:}
  \begin{equation}
    \label{eq:S-CFT-L}
    C^\mu(x) = \matel{0}{{\cal O}_1(x) {\cal O}^\mu_3(0)}{0}  \sim
    \begin{cases}
      x^\mu (x^2)^{-(\alpha+1/2)}\phantom{0}  \quad \text{SFT} \\[0.1cm]
      0\phantom{x^\mu (x^2)^{-(\alpha+1/2)}} \quad   \text{CFT}   \end{cases} \;, 
  \end{equation}
  with $\alpha \equiv (\Delta^*_{{\cal O}_1}+\Delta^*_{{\cal O}_3})/2$. 
  Eq.~\eqref{eq:S-CFT-L} follows from the investigations in \cite{Osborn:1993cr}.
\end{enumerate}
The equations above state that, in order to have a non-vanishing correlator, the
scaling dimension as well as the spin structure of the two operators
in question have to be identical~\cite{Osborn:1993cr}.  Let us add
that it is the \emph{local} nature of the special conformal
transformations which is responsible for the selection rules quoted
above.  These transformations are precisely the difference between the
symmetries of a CFT and a SFT.
 
Using  Eq.~\eqref{eq:Gtof} we get:
\begin{equation}
\label{eq:scale-f}
F(t,y) \stackrel{t \to 0} {\rightarrow}  \begin{cases}
\text{constant}\phantom{0}  \quad \text{SFT} \\[0.1cm]
0\phantom{\text{constant} } \quad   \text{CFT}   \end{cases} \;,
\end{equation}
for $y = \Lambda/m_0 = \Lambda \sqrt{x^2}/b $ such that the system is suitably close to the fixed point.  More precisely for fixed $x^2$ and $\Lambda$, $y$ (or $b$) has to be such that the system is close to the fixed point. 
In general we expect the  constant to be finite with the possible caveat that the correlator, 
which is generally not a physical observable, is affected by IR-divergences. 

This criterion is unfortunately of limited use for standard gauge
theories.  In recent years the understanding has
emerged~\cite{Fortin:2011sz} that limit cycles are the only
possibility, for four-dimensional unitary quantum field theories to be
scale but not conformal invariant. On the other hand limit cycles have
only been found in theories with flavour dependent couplings, aka
Yukawa terms~\cite{Fortin:2011sz}. These couplings are absent in the
gauge theories currently studied on the lattice, and therefore it
would seem that IR-conformal theories are indeed IR-conformal and not
just IR-scale-invariant.  Let us add to this end that, currently, the
only logical possibility for scale invariant theories to exist is if
the theories can evade the strong version of the a-theorem at the
non-perturbative level~\cite{Fortin:2012hn}, as the
latter has been shown to be valid in perturbation theory some time
ago~\cite{Jack:1990eb}.

\subsection{CFT (IR-conformal) vs confining theory}
\label{sec:CFTvsConf}

In the previous paragraph we discussed a possible recipe for
discerning theories that have only one (flavour independent) coupling,
and that are CFTs and not SFTs.  The selection rules can also be
useful in distinguishing CFTs (IR-conformal) from confining
theories. In the following we shall assume that in a gauge theory
without IR fixed points, chiral symmetry breaking and confinement
occur together.

For that purpose, let us analyse the dimension, and the spin selection
rules for a number of example operators. We consider the case of a)
quasi-primary operators from the viewpoint of the CFT candidate
theory, b) whose correlation function does not vanish by virtue of
non-CFT selection rules such as parity symmetry for example.
\begin{enumerate}
\item \emph{Scaling dimension:}\footnote{The actual implementation on
    the lattice might still be non-straightforward as a the gluon
    field strength tensor is known to mix with $m \bar qq$.  Yet in
    the limit this mixing would disappear. A more delicate issue is
    the mixing with the identity, which corresponds to the
    disconnected part of the correlator. Whereas in dimensional
    regularisation the mixing occurs  with $m^4 \mathbf{1}$ only,
    which is of no problem for the same reason as above, in lattice
    cut-off regularisation terms of the form $m^2 \Lambda^2
    \mathbf{1}$ and $ \Lambda^4 \mathbf{1}$ (with $\Lambda = 1/a$ with
    $a$ being the lattice spacing) are expected to occur. 
    Whereas  both have 
    hitherto prohibited a clean extraction of the gluon
    condensate for instance it is the latter which seems to pose a problem for the case the discussed above.}
  \begin{alignat}{2}
    \label{eq:proposal1}
    &{\cal O}_1 = \frac{1}{g^2} G^2 \;, \quad &\Delta^*_{G^2} =&  4  \;, \nonumber  \\[0.1cm]
    &{\cal O}_2 = \bar q q   \;, \quad &\Delta^*_{\bar q q } =&  3 -\gamma_m^* \; .
  \end{alignat}
  We note that the correlator \eqref{eq:S-CFT-G} with
  \eqref{eq:proposal1} vanishes to all order in perturbation theory in
  the massless limit as the gauge theory Lagrangian is even under $q \to
  \gamma_5 q$ and $m \to -m$ whereas the correlator is odd (since $\bar
  q q \to - \bar q q$ and $G^2 \to G^2$). Thus the correlator probes the
  non-perturbatve regime or more precisely chiral symmetry breaking through
  $\vev{\bar qq } \neq 0$.
  
\item \emph{Spin:}
  \begin{alignat}{2}
    \label{eq:proposal2}
    &{\cal O}_1 =  P^a_5 = \bar q i \gamma_5 t^a q \;, 
    \quad & &\Delta^*_{P^a_5} =  3 - \gamma_m^*  \;, \nonumber  \\[0.1cm]
    &{\cal O}^\mu_3=  A^{a\,\mu} = \bar q i \gamma^\mu \gamma_5 t^a q  
    \;, \quad & &\Delta^*_{A^{a\,\mu}} =  3 \;,
  \end{alignat}
  where $t^a$ is a $SU(N_f)$ representation matrix acting on flavour
  space.  For the same reason as above the correlator \eqref{eq:S-CFT-L}
  with \eqref{eq:proposal2} vanishes to all orders in perturbation
  theory in the massless limit.  It is, however,  non-vanishing in the
  theory with chiral symmetry breaking since the pion couples to both
  currents:
  \begin{equation}
    \matel{0}{A_\mu^a(0)}{\pi^b} = \delta^{ab} i  f_\pi p_\mu \;, \qquad  
    \matel{0}{P^a(0)}{\pi^b} = \delta^{ab} g_\pi \;, \qquad g_\pi = \frac{f_\pi m_\pi^2}{2m} \;,
  \end{equation}
  where it is noted in particular that $g_\pi$ is finite and non-vanishing in the chiral limit 
  for $m_\pi^2 \sim m$ in a chirally broken phase. Conversely $f_\pi$ is only non-vanishing 
  if $m_\pi \sim m$ at least which is the case for the Goldstone bosons only. 
  It seems worthwhile to elaborate a bit further on this point. The correlation function 
  \eqref{eq:S-CFT-L} assumes the following form in the chirally broken phase,
  \begin{equation}
    C^\mu(x) = \matel{0}{{\cal O}_1(x) {\cal O}^\mu_3(0)}{0}  = 
    \underbrace{\int \frac{d^4 p}{(2\pi)^4} \frac{i p_\mu f_\pi g_\pi
      }{p^2 + m_\pi^2} e^{i p \cdot x}}_{ \equiv C_0^\mu(x)} + {\cal O}(m) \;,
  \end{equation}
  with all other contributions to the spectrum vanishing in the chiral
  limit\footnote{Multiparticle pion states also come with zero invariant
    mass but at the same time have zero phase space and therefore vanish
    in the limit $m \to 0$.}.  More precisely 
  \begin{eqnarray} 
  \label{eq:C0}
    C_0^\mu(x)
    &=&
    f_\pi g_\pi \partial_\mu   \int \frac{d^4 p}{(2\pi)^4}\frac{e^{i p
        \cdot x} }{p^2 + m_\pi^2}  
    = f_\pi g_\pi \partial_\mu \frac{m_\pi K_1(m_\pi x)}{(2\pi)^2 x}  \nonumber \\[0.1cm]
    & =& f_\pi g_\pi \partial_\mu \left( \frac{1}{x^2} + {\cal O}(m_\pi
      \ln x) \right) = \vev{\bar qq} \frac{2 x^\mu}{x^4} + {\cal O}(m)
    \;,
  \end{eqnarray}
  where $x = \sqrt{x^2}$, for odd powers of $x$ and in the last equality
  we have made use of the Gell-Mann Oakes Renner (GMOR) relation $ f_\pi ^2 m_\pi^2 = - 2m
  \vev{\bar qq}$.
\end{enumerate}

Let us briefly comment on the scaling dimensions of the operators
quoted in Eqs.~(\ref{eq:proposal1},\ref{eq:proposal2}) which were
already exploited in our previous work
\cite{DelDebbio:2010ze,DelDebbio:2010jy}. The scaling dimension of the
gluon field strength tensor is four as it appears in the trace anomaly
which is related to the physical mass. The scaling dimension of quark
condensate times the mass is four, $\Delta^*_{\bar qq} + (1+\gamma_m^*)
= 4$, for the same reason and therefore $\Delta^*_{\bar q q } = 3
-\gamma_m^* $.  The scaling
dimension of $A^{a\,\mu}$ is three because it is a partially conserved
current affected only by explicit breaking. The scaling dimension of $
P^a_5 $ can be obtained from Ward Identities as presented in appendix
B.1 of Ref.~\cite{DelDebbio:2010ze}. It would seem worthwhile to point
out that the operators quoted  in 
Eqs.~(\ref{eq:proposal1},\ref{eq:proposal2}) are of the
(quasi)-primary type as the non-primary operators derive from the
latter through derivatives.

Finally in essence we get, as a replacement of Eq.~\eqref{eq:scale-f}
for the case at hand,
\begin{equation}
  \label{eq:scale-f2}
  F(t,y) \stackrel{t \to 0} {\rightarrow}  \begin{cases}
    \neq 0 \phantom{0}  \quad \text{confining} \\[0.1cm]
    0\phantom{\neq 0  } \quad   \text{CFT}   \end{cases} \;,
\end{equation}
for $y$ such that the system is suitably close to the fixed point as previously discussed.
We note that $F(t,y)$ is known explicitly \eqref{eq:C0}
for the second example considered.

\subsection{Comments on finite volume effects in lattice simulations}

\begin{figure}[h]
  \centering
  \includegraphics[width=3.5in]{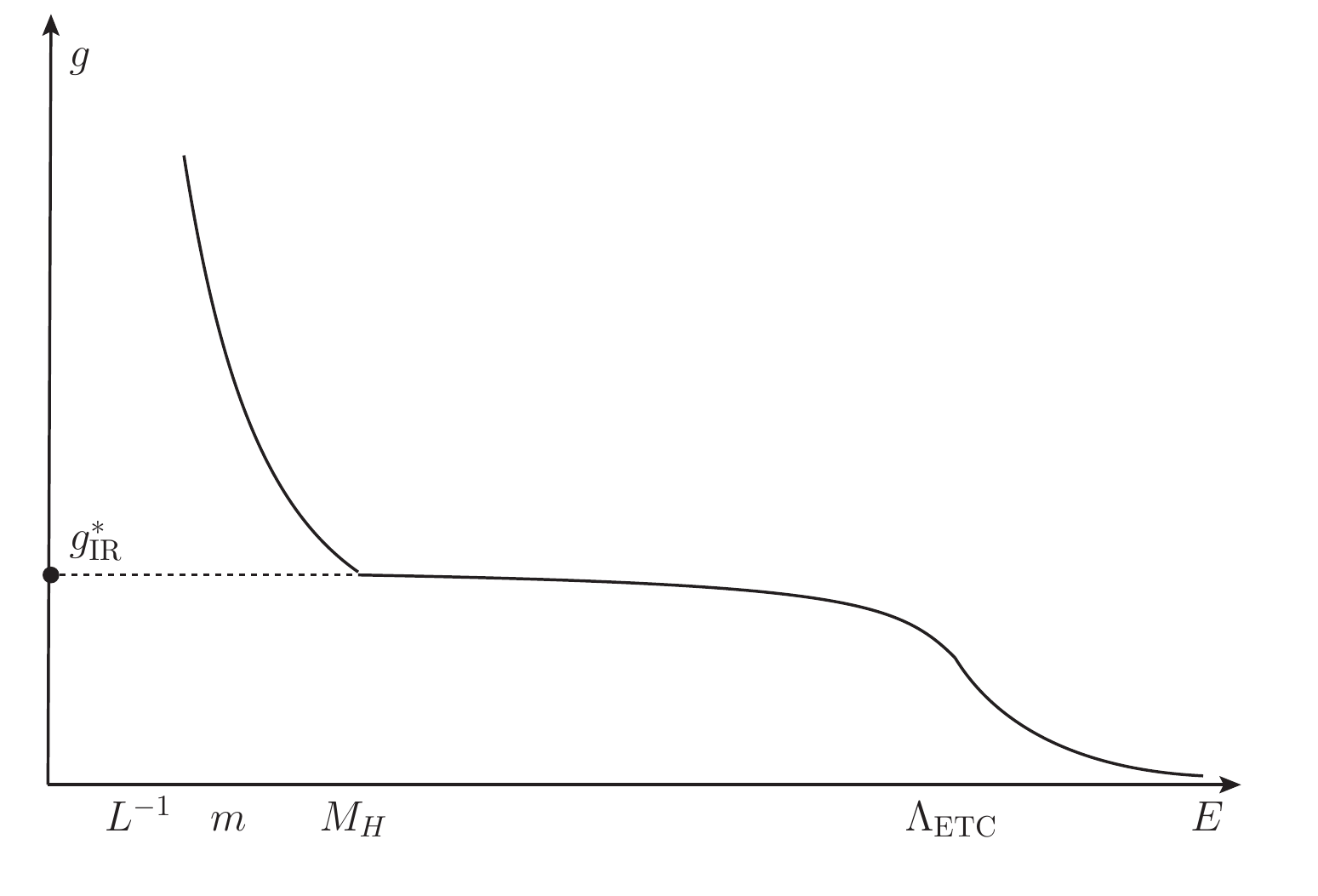} 
  \caption{\small Sketch of the RG flow for an IR-conformal gauge
    theory. At high energies the theory is asymptotically free, and at
    lower energies it reaches a fixed point $g_{\rm IR}^*$. The mass
    parameter $m$, or equivalently the the mass scale $M_H > m$ (of
    the hadronic bound states), drives the theory away from the
    FP. This is the picture that heuristic computations of the quark
    condensate suggest e.g. \cite{Sannino:2008nv,DelDebbio:2010ze}.
    Note the inverse of the lattice box size has to be significantly
    smaller than $M_H$ in order for FSE effects to be under
    control. More precisely as long as $L M_H \gg 1$, FSE are of the
    order $\exp(- M_H L)$.  If this condition is not met the effects
    are power-like with known exponents e.g.~\cite{Kuti}.  We
    should point out that we have not attempted to indicate the effect on 
    the coupling 
    of the actual value of $L$ on the curve on the graph. }
\label{fig:coupling}
\end{figure}

Eqs.~(\ref{eq:scale-G}), and (\ref{eq:scale-f}) show that the behaviour
of $F(t,y) \to 0$, as $t \to 0$, provides the possibility to
distinguish between a CFT and an SFT or a confining theory.  In
practice one would keep $x$ fixed and study the behaviour of the
correlator for $m \to 0$.  In lattice simulations one would need to
work for given values of $m$ with sufficiently large volumes, $L \gg {M_H}^{-1}$, as the limits $m \to 0$ and $V \equiv L^4 \to \infty $
are known not to commute.  Further comments can be inferred from
Fig.~\ref{fig:coupling} where the relative scales are sketched against
a typical behaviour of a (gauge) coupling.  In regard to this figure we would
like to draw the readers attention to the fact that the actual value
of the coupling is scheme dependent, whereas the question of whether
there is a fixed point or not is scheme independent as it shows up in
physical measurable quantities in terms of scaling laws.

\section{First order correction to the fixed point}
\label{sec:first}

We have already discussed the scaling of field correlators as a
function of the mass $m$ for $g=g^*$. When the coupling is not tuned
to its critical value, scaling corrections appear. In this section we
compute these corrections at first order in the $\delta g \equiv g-g^*$.
In section~\ref{sec:linear} we introduce the notation and discuss the
linearized RG equations. In section~\ref{sec:sc} we compute the
scaling corrections to field correlators of local operators.  In
section~\ref{sec:scM} we apply these results and compute the scaling
corrections to the hadronic masses first by using the trace anomaly,
and then by analysing the mass correlator. Furthermore we show that
the two expressions for the scaling corrections are equal by using an
extension of the Feynman-Hellmann theorem \cite{gluon-relations}. 

\subsection{Linearisation around the IR fixed-point}
\label{sec:linear}

We assume that the bare couplings at the cut-off scale $g$ and $\hat
m$, which correspond to the point $I$ in Fig.~\ref{fig:3-types}(left)
with the identification $(g,g') =(g,\hat m)$, are chosen such that the
system is on a trajectory that is close to the fixed point.  We are
going to linearize the RG flow equations in the deviations from the
fixed point, that is to say in the variable $\delta g \equiv g-g^*$,
where we use the notation $g^* = g^*_{\rm IR}$ throughout this
section. We shall comment on the aspects of this expansion at the end
of section \ref{sec:discuss}.  For the beta function, the mass anomalous
dimension $\gamma_m$, and the anomalous dimension matrix $\gamma_{ij}
\equiv (\gamma_O)_{ij} $\footnote{In statistical mechanics the
  anomalous dimension of operators are often denoted by the symbol
  $\eta$ rather than $\gamma$ in order to distinguish it from
  anomalous dimensions of parameters such as the mass for instance.}
of a generic set of operators $\{ O_i \}$ that mix under the RG-flow,
we may linearize the system around the IRFP as follows:
\begin{figure}[h]
  \centering
  \includegraphics[width=3.5in]{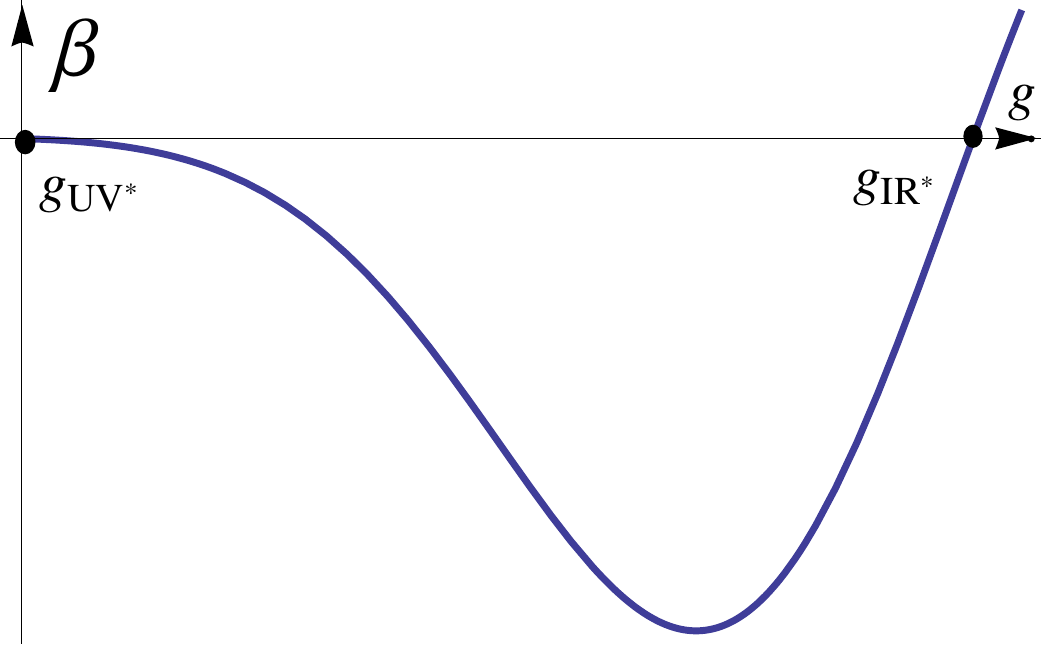} 
 \caption{\small Sketch of the $\beta$ function in terms of the coupling for a system 
 exhibiting asymptotic freedom $g^*_{\rm UV} = 0$ and a non-trivial IR fixed point
 at some value $g^*_{\rm IR} > 0$. In section \ref{sec:first} a system in the vicinity 
 of the IR fixed point is considered as indicated in the figure.}
\label{fig:one}
\end{figure}
\begin{alignat}{2}
  \label{eq:ansatz}
  \beta &=   \beta_1 \delta g  + {\cal O}( \delta g^2) \;, 
  \qquad \qquad \quad   & &  \delta g \equiv g-g^*   \;,
  \nonumber \\[0.1cm]
  \gamma_m &= \gamma_m^* +  \gamma_m^{(1)} \delta g  +
  {\cal O}( \delta g^2) \;, & & \nonumber \\[0.1cm]
  \gamma_{ij} &= \gamma^*_{ij} +  \gamma_{ij}^{(1)} \delta g  + 
  {\cal O}( \delta g^2)  \;, & & ( \gamma_{ij} \equiv (\gamma_O)_{ij} ) \;.
\end{alignat}
We have verified that in a mass independent scheme $\beta_1$ is
universal (scheme independent) whereas $\gamma^{(1)}_{m/ij}$ are
not. We remind the reader that the anomalous dimensions associated
with gauge invariant operators (such as $\gamma_m^*$) are universal.
When working with renormalized quantities we shall choose notation
accordingly.  The behaviour of the beta function as a function of the
coupling is illustrated in Fig.~\ref{fig:one}; $\beta_1$ corresponds
to the slope where the curve crosses the IR fixed point.  We note that
for the beta function described in Fig.~\ref{fig:one}, the coefficient
$\beta_1$ is positive as there are no further zeros between $g=0$ and
$g=g^*$.  The beta function equation is easily integrated to that
order,
\begin{equation}
  \label{eq:gt}
  \beta(g) =   \Lambda \frac{d}{d \Lambda} (\delta g)  = 
  \beta_1 \delta g  + {\cal O}( \delta g^2)  \quad \Rightarrow \quad 
  \delta g(\Lambda) =   \delta g(\Lambda_0) \left( \frac{\Lambda}{\Lambda_0} \right)^{\beta_1} \;,
\end{equation}
where $\Lambda$ is a UV cut-off as will become clear in the next subsection.

\subsection{Scaling corrections to correlators}
\label{sec:sc}

We shall use the language of Wilsonian renormalization group for which
the theory is defined at some fixed UV cut-off $\Lambda_{\rm UV}
\equiv \Lambda$\footnote{In the context of a lattice field theory the
  lattice spacing $a$ is related to the UV cut-off as $a =
  \Lambda_{\rm UV}^{-1}$.}.  Let us consider a correlation function
$O_i(g,m,\Lambda)$, as a function of the bare parameters $g,m$, and
the UV cut-off $\Lambda$. We shall denote by $Z_{ij}$ the matrix that
describes the mixing of $O_i$ under renormalization. $O_i$ satisfies an
RG equation (also known as 't Hooft-Weinberg or Callan-Symanzik
equations), e.g. Ref.~\cite{ZinnJustin:2002ru},
\begin{equation}
  \label{eq:RGWilson}
  \left( 
    \Lambda \frac{\partial}{\partial \Lambda}  \delta_{ij} +  \beta(g)
    \frac{\partial}{\partial g} \delta_{ij}
    - \gamma_m  m \frac{\partial}{\partial  m} \delta_{ij} - \gamma_{ij}
    \right) O_j(g, m,\Lambda) = 0 \;,
\end{equation} 
where summation over $j$ is implied and 
\begin{equation}
  \beta(g) =   \Lambda \frac{d}{d \Lambda} g \;, \quad 
  \gamma_m = -   \Lambda \frac{d}{d \Lambda} \ln m \;,  \quad
  \gamma_{ij} =   
  \Lambda \frac{d}{d \Lambda} \ln Z_{ij} \;.
\end{equation}
We now wish to reformulate the theory using 
a different UV cut-off $\Lambda'_{\rm UV} \equiv \Lambda'$
\begin{equation}
\frac{\Lambda}{\Lambda'} = b  \;,
\end{equation}
where the parameter $b$ has the interpretation of a blocking factor,
and $b > 1$ if high-energy modes are to be integrated out. The formal
solution to Eq.~\eqref{eq:RGWilson} is given by:
\begin{equation}
  \label{eq:ansatz}
  O_i(g,m,\Lambda) = Z_{ij}^{-1}(b) O_j(g(b), m(b), \Lambda/b)
  \;,
\end{equation}
where 
\begin{alignat}{3}
  \label{eq:RC1}
  & \frac{d}{d \ln b} \ln Z_{ij}(b) \; &=& - \gamma_{ij}(g(b)),
  \qquad  \quad  & & Z(1)=1 \;, \nonumber  \\
  & \frac{d}{d \ln b} g(b) &=& -  \beta(g(b)),
  & & g(1)=g  \;, \nonumber  \\
  & \frac{d}{d \ln b} \ln m(b) &=& \phantom{-} \gamma_m (g(b)),
  & & m(1)=m \, .
\end{alignat}
We assume here that we are working in a \emph{mass independent
  scheme}, and therefore the beta function and the anomalous
dimensions only depend on the gauge coupling $g$.  The solution
(\ref{eq:ansatz},\ref{eq:RC1}) is known by the name of the method of
characteristics, see e.g. Ref.~\cite{ZinnJustin:2002ru}.  Assuming the
fixed point is in the linear regime~\eqref{eq:gt}, the three equations
above can be solved to order ${\cal O}(\delta g)$:
\begin{alignat}{2}
  \label{eq:chara_sol}
  &   g(b)  \; & =& \; g^* + \delta g(b)  = g^*  +  \delta    g \, b^{-\beta_1}\, ,   \nonumber  \\
  &  m(b) &=& \;m b^{\gamma^*_m}
  \exp\left[-\frac{\gamma_m^{(1)}}{\beta_1}
    \delta g f(b)  \right]\, , \nonumber  \\
  & Z_{ij}(b) &=&\;  \exp\left[ \gamma^* \ln b  - \frac{\gamma^{(1)}}{\beta_1}
    \delta g f(b) \right]_{ij} \;,
\end{alignat}
where we have introduced the notation 
\begin{equation}
  f(b) \equiv b^{-\beta_1}  -1   \;, 
\end{equation}
which parameterizes the distance from the initial point in blocking
space.  Eq.~\eqref{eq:ansatz} may be written using the relation
\eqref{eq:chara_sol} as:
\begin{eqnarray}
  \label{eq:NLO}
  O_i(g,m,\Lambda) &=&  Z_{ij}(b) ^{-1} O_j(g(b), m(b), \Lambda/b) \nonumber  \\
  &=& 
  \exp\left[ - \gamma^* \ln b + \frac{\gamma^{(1)}}{\beta_1}
    \delta g f(b) \right]_{ij} 
  O_j( g(b), m b^{\gamma_m^*} \exp\left[- \frac{\gamma_m^{(1)}}{\beta_1}
    \delta g f(b)  \right] , \Lambda/b)  
  \nonumber \\
  &=&  \exp\left[ - \Delta^* \ln b + \frac{\gamma^{(1)}}{\beta_1}
    \delta g f(b) \right]_{ij} 
  O_j( g(b),m b^{-(1+\gamma_m^*)} \exp\left[- \frac{\gamma_m^{(1)}}{\beta_1}
    \delta g f(b)  \right] , \Lambda)  \nonumber \;,
\end{eqnarray}
where in the last equality we have rescaled all dimensionful
quantities by a factor $b$. The matrix $\Delta_{ij}=d_i \delta_{ij} +
\gamma_{ij}$, where $d_i$ is the classical dimension of $O_i$, yields
the scaling dimensions of the operators. In order to get the $\delta g$
corrections we need to expand in that variable.  In our opinion this
is best done from the expression in the the second line of the
equation above. The last step can be done after the expansion for each
individual term. Furthermore, in order to avoid path ordering in
coupling space, we shall assume that $\gamma_{ij}^*$ is diagonal. The
corrections are parameterized as follows,
\begin{equation}
  \label{eq:NLO_taylor}
  O_i(g,m,\Lambda) = b^{-\gamma_{ii}^*}  
  \left(    [O_i]^* + \delta g \,  O^{(1)}_i +   {\cal O}(  \delta g^2) \right) \;,
\end{equation}
where 
\begin{equation}
  O^{(1)}_{i}  =   \left(    
    \frac{\gamma^{(1)}_{ii} }{\beta_1}
    [O_i]^*f(b) - 
    \frac{\gamma_m^{(1)}}{\beta_1}  m^*[O_i]^*_{,m} f(b)  +
    [O_i]^*_{,g}  b^{-\beta_1}      
  \right) 
\end{equation}
and
\begin{eqnarray}
\label{eq:star}
  [O_i]^*&=&  O_i(g(b) , m(b) ,\Lambda/b)|_{\delta g =0}   \;, \nonumber \\[0.1cm]
  [O_i]^*_{,m} &=&  \frac{\partial}{\partial m(b) } 
  O_i(g(b) , m(b) ,\Lambda/b)|_{\delta g =0}    \;, \nonumber \\[0.1cm]
  [O_i]^*_{,g}  &=&  \frac{\partial}{\partial g(b)} 
  O_i(g(b) , m(b) ,\Lambda/b)|_{\delta g =0} \nonumber  \;, \\[0.1cm]
  m^*  &=& m(b)|_{\delta g  =  0 }  \;, \quad ( g^* = g(b)|_{\delta g  =
    0 })  \, .
\end{eqnarray}
We wish to emphasise that, when $g^*$ is tuned to the fixed point
coupling, $m^* = m b^{\gamma^*_m}$ corresponds to the leading scaling
of the mass at the fixed point.  

The scaling corrections as a function of $m$ can be made explicit by
rescaling all dimensionful quantities by the appropriate power of $b$
in the last step in Eq.~\eqref{eq:NLO}, and then using the
arbitrariness of $b > 1$ to impose:
\begin{equation}
  mb^{-(1+\gamma_m^*)} = m_0 \quad \Rightarrow b^{-1} =  
  \hat m^{1/(1+\gamma_m^*)} \;, \quad  \hat m  \equiv \frac{m}{m_0} \;.
\end{equation}
As a result, we obtain a scaling formula that includes the scaling
corrections at first order in $\delta g$: 
\begin{eqnarray}
  \label{eq:schematic}
  O_i(g,m,\Lambda)  &=& \hat m^{\frac{\Delta_{ii}}{1+\gamma_m^*} } [O_i]^*
  \left( 1 + \delta g ( A + B \,\hat m^{\frac{\beta_1}{1+\gamma_m^*}}
    )\right)  +   {\cal O}(  \delta g^2)  \;,
\end{eqnarray} 
with 
\begin{eqnarray}
\label{eq:AB}
  A &=&  \left\{  - \frac{\gamma^{(1)}_{ij} }{\beta_1}   + 
    \frac{\gamma_m^{(1)}}{\beta_1}  m^* \frac{[O_i]^*_{,m}}{[O_i]^*}  \right\}_b \nonumber \\
  B &=&  \left\{   +  \frac{\gamma^{(1)}_{ij} }{\beta_1} - 
    \frac{\gamma_m^{(1)}}{\beta_1}  m^* \frac{[O_i]^*_{,m}}{[O_i]^*} +
    \frac{[O_i]^*_{,g}}{[O_i]^*}
 \right\}_b  \;,
\end{eqnarray}
where the curly brackets with a $b$ superscript indicate that all
physical units are to be scaled by $b$, e.g. $\{[O_i^*] \}_b \stackrel{\eqref{eq:star}}{=} 
\{  O_i(g(b) , m(b) ,\Lambda/b)|_{\delta g =0}\}_b \to b^{d_O}   
O_i(g(b) , b m(b) ,\Lambda)|_{\delta g =0}$. It is interesting to note
that the scaling corrections above simplify when $\beta_1 \to 0$:
\begin{equation}
  \label{eq:BZlimit}
  ( A + B \,\hat m^{\frac{\beta_1}{1+\gamma_m^*}} ) \to
    \frac{[O_i]^*_{,g}}{[O_i]^*} + O(\beta_1)\, . 
\end{equation}
This situation is expected to be realised at the lower edge of the
conformal window in the Banks-Zaks limit. 

\subsubsection{Discussion of scaling corrections}
\label{sec:discuss}

The expression \eqref{eq:NLO_taylor} yields the corrections to scaling
for small fermion mass $\hat m$, while the irrelevant coupling $g$ is
at a distance $\delta g$ from the fixed point. Clearly when $\delta g$
vanishes, so do the scaling violations.  We note that for fixed
initial value $g$, $\delta g \equiv g - g^*$ is proportional to the value of
the IR fixed point coupling $g^*$, as can be inferred from
Fig.~\ref{fig:3-types}.  The linear approximation discussed
here becomes therefore less reliable if the IR fixed point is at strong coupling
coupling, unless $g$ is tuned to reduce the size of $\delta g$. Note
that for large $\delta g$ the linear corrections tend to grow. This
can be compensated by going to smaller initial masses $\hat m$ since
the first order (relative) scaling corrections are determined by the
combination $B \delta g\, \hat m^{\beta_1/(1+\gamma_m^*)}$.
  
We would like to add an important point concerning the size of the
corrections at the lower edge of the conformal window. For a strong
coupling fixed point, one would expect large values of
$\gamma_{ij}^{(1)}$, as well as $\gamma_{m}^{(1)}$, whereas the value
of $\beta_1$ is expected to be small as the fixed point is to be lost
which in turn is consistent with $g^*$ being large. Moreover, unless the
bare coupling $g$ is fine-tuned, one can expect to have rather large
values of $\delta g$, driven by our ignorance in guessing the exact
location of the fixed point. Thus in summary the precoefficient $B
\delta g$ should be expected to be large at the lower edge of the
conformal window.  On the other hand the exponent
$\beta_1/(1+\gamma_m^*)$ is then small and leads to a suppression. In
the previous statements large and small are meant relative to the
region away from the lower edge of the conformal window.  Which of the
two counteracting effects dominates is unclear a priori but the
argument suggests that it is important to go to small masses $\hat m$
at the lower edge of the conformal window to suppress potentially
large scaling corrections. This is of practical importance as many of
the lattice simulations have been performed precisely at the lower
edge of the conformal window in search of a theory of walking
technicolor.

The signs of $A$ and $B$, in Eq.~\eqref{eq:schematic}, are determined by the 
dynamics. Since in general we cannot make statements about the derivative of the operators the sign of $A$ and $B$ are thus not known a priori. 
This is somewhat different for the hadronic masses that is to say for the operators 
$Q$ and $G$ which is what we are going to exploit in the next section.

\subsection{Scaling corrections to the mass formula}
\label{sec:scM}
We shall first introduce some notation and justify the formulae needed
for the comparison of the two derivations of the scaling corrections to
the hadronic mass in subsection~\ref{sec:pathways}.

\subsubsection{Preliminary formulae }
The following notation, 
\begin{equation}
  \label{eq:X}
    \vev{X}_{\EH}   \equiv     \matel{H(E,\vec{p})}{X}{H(E,\vec{p})}_c  \;,
\end{equation}
shall prove convenient throughout this section. The subscript $c$
denotes the connected part of the matrix element, while $|H(E,\vec{p})
\rangle$ is a physical state with definite spatial momentum and
energy and $X$ is a (local) operator. Above we have explicitly
indicated the energy dependence of the hadronic state $H$ which we
occasionally suppress in the remaining part of this work.  Note that
the disconnected part of the correlator is related to the vacuum
energy, that is to say the cosmological constant.  As usual the Lorentz invariant
state normalisation is given by:
\begin{equation}
  \label{eq:normal}
  \vev{H(E',\vec{p'})|H(E,\vec{p})} = 2 E(\vec{p})   (2 \pi)^3 \delta^{(3)}(\vec{p}-\vec{p'}) \;. 
\end{equation}

The expectation value of the energy momentum tensor $T_{\mu\nu}$ in a
single-particle state is:
\begin{equation}
  \label{eq:Tmunu}
  \vev{T_{\mu\nu}}_{E_H} 
  = 2 p_{\mu}  p_{\nu}\, ,
\end{equation}
where $p_0 = E$.  In order to keep a compact notation we are going to extend the notation 
\eqref{eq:X} for two specific matrix elements to:
\begin{eqnarray}
  \label{eq:GQ}
  Q_{\EH}&\equiv&   N_f m \vev{ \bar{q}q}_{E_H}  \;,  \nonumber \\[0.1cm] 
  G_{\EH} &\equiv&  
  \vev{ \frac{1}{g^2} G^2}_{E_H} \, .
\end{eqnarray}
We remind the reader that the notation \eqref{eq:X} 
refers to the connected part of the matrix element only.

For the discussion in this section it is convenient to use
renormalized quantities. Accordingly we denote the renormalized
couplings by $\bar{g}$ and $\bar{m}$, and the matrix elements of the
renormalized operators $\bar{G}_{E_H}$ and $\bar{Q}_{E_H}$
respectively. The renormalized coupling $\bar{g}$ is defined as:
\begin{equation}
  \label{eq:Zg}
  g = Z_g(g) \bar{g}\, .
\end{equation}
En passant we note that physical quantities such as the energy momentum tensor and
thus the hadronic mass do not renormalize (i.e. $T_{\mu \nu} = \R{T}_{\mu\nu}$).
In the neighbourhood of the IRFP, the renormalization constant is expanded similar to\eqref{eq:ansatz} as,
\begin{equation}
  \label{eq:Zgstar}
  Z_g(g) = Z_g^* + Z_g^{(1)} \delta g + \mathcal{O}(\delta g^2)\, ,
\end{equation}
which implies: 
\begin{equation}
  \label{eq:geqs}
  \delta g = (Z_g^* + g Z_g^{(1)}) \delta \bar{g}  + \mathcal{O}(\delta g^2) \, ,~~~~g \frac{\partial}{\partial g} =
  \kappa \R{g} \frac{\partial}{\partial \bar{g}}  \, , \quad \kappa = \left( 1 - \frac{Z_g^{(1)}}{Z_g^*} \right) + \mathcal{O}(\delta g^2) \;.
\end{equation}
The trace anomaly can be written in terms of the
renormalized quantities as \cite{EMTtrace}:
\begin{equation}
  \label{eq:master_b}
  2 M_H^2 =  \left( \frac{\R{\beta}}{2 \bar{g}}\right)
  \R{G}_{M_H} + (1 +\R{ \gamma}_m ) \R{Q}_{M_H} \, , 
\end{equation}
and is an RG-invariant.  More precisely since $\bar{Q}_{\EH}$ is an
RG invariant, $\R{\beta}/(2 \bar{g}) \R{G}_{\EH} + \R{ \gamma}_m
\bar{Q}_{\EH}$ inherits this property by virtue of Eq.~\eqref{eq:master_b}. 
This entails that $G_{\EH} \neq \R{G}_{\EH} $.
In identifying the two computation the following relations are of importance:
\begin{equation}
\label{eq:FH}
\R{m} \frac{\partial}{\partial \R{m}}   E_H^2 = \R{Q}_{E_H} \;, \qquad  \quad 
\R{g} \frac{\partial}{\partial \R{g}}   E_H^2 = - \frac{1}{2} \R{G}_{E_H}  \;.
\end{equation}
The first relation is a straightforward application of the Feynman-Hellmann theorem 
and is widely used, as for instance in our previous work \cite{DelDebbio:2010jy}.
The second relation is akin to a Feynman-Hellmann relation. It has been derived 
in \cite{gluon-relations}
through an RGE, the trace anomaly  \eqref{eq:master_b} as well as the first relation in \eqref{eq:FH}.
Later it was rederived and checked in a few exactly solvable models in \cite{hamilton}.

\subsubsection{Two pathways to mass-scaling corrections}
\label{sec:pathways}

Let us now compute the corrections to scaling in two different ways by
using results from the previous section: the corrections are obtained
up to order $\delta \bar{g} \equiv (\bar{g} - \bar{g}^*) $ and the
symbol $\delta$ on other quantities denotes the linear variation in
the $\delta \bar{g}$ variable. Recall that
$\beta_1 = \R{\beta_1}$ (at least in a mass independent scheme) 
and $\gamma_m^{(1)} \neq \R{\gamma}_m^{(1)}$ in general and we shall therefore use notation accordingly.

\begin{enumerate}
\item First we compute $\delta( 2M_H^2)$ directly from the RG scaling
  formulae~\eqref{eq:schematic} for renormalized quantities, 
  combined with the relation \eqref{eq:FH}:
  \begin{eqnarray}
    \label{eq:direct}
    \delta(2M_H^2) &=& \delta \R{g} \left(  [ 2M_H^2]^*_{,\R{g}} b^{-\beta_1}  
    -  \frac{\R{\gamma}_m^{(1)}}{\beta_1} \R{m}^* [ 2M_H^2]^*_{,\R{m}} f(b)   \right) 
    + {\cal O}(\delta \R{g}^2)  \nonumber \\
    &\stackrel{\eqref{eq:FH}}{=}&\delta \R{g} 
    \left(  - \frac{1}{\R{g}^*} [\R{G}_{M_H}]^* b^{-\beta_1}   - 
    2 \frac{\R{\gamma}_m^{(1)}}{\beta_1} [\R{Q}_{M_H}]^* f(b)
    \right) 
    + {\cal O}(\delta \R{g}^2) \;.
  \end{eqnarray}
\item Second we  compute $\delta( 2M_H^2)$ through the
  trace anomaly~\eqref{eq:master_b}:
  \begin{eqnarray}
    \label{eq:indirect}
    \delta(2M_H^2) =\delta \R{g} b^{-\beta_1} \left( 
      \frac{\beta_1}{2 \R{g}^*}
      [\R{G}_{M_H}]^* + 
      \R{\gamma}_m^{(1)} [\R{Q}_{M_H}]^* 
      \right) +  \left(1+ \gamma_m^*\right) \delta \R{Q}_{M_H} \;,
  \end{eqnarray}
  which necessitates the computation of $\delta \R{Q}_{M_H}$. The latter
  is given by Eq.~\eqref{eq:schematic}:
  \begin{eqnarray}
    \label{eq:dQ}
    \delta \R{Q}_{M_H}  =  \delta \R{g} \left(  [ \R{Q}_{M_H}]^*_{,\R{g}} b^{-\beta_1}
    - \frac{\R{\gamma}_m^{(1)}}{\beta_1}\R{m}^*  [ \R{Q}_{M_H}]^*_{,\R{m}} f(b)   \right) \;.
  \end{eqnarray}
\end{enumerate}
The expressions~\eqref{eq:direct}, and~\eqref{eq:indirect},
\eqref{eq:dQ} yield the scaling corrections as a function of $\delta
g$ and the mass $m$. These expressions all have the same scaling
exponents, yet it is not clear from these formulae that the
corresponding prefactors are equal.  To compare the prefactors we
ought to compute $[\R{Q}_{M_H}]^*_{,\R{g}}$ and $[\R{Q}_{M_H}]^*_{,\R{m}}$
to leading order in $\delta \R{g}$. The latter is simply given by the leading order scaling 
\eqref{eq:master}
\begin{equation}
  \R{m}^*  [\R{Q}_{M_H}]^*_{,\R{m}} = 
  \frac{2}{1+\gamma_m^*} [\R{Q}_{M_H}]^*  + {\cal O}(\delta \R{g}) \;,
\end{equation}
up to corrections which are beyond the aimed accuracy.
The computation of $[\R{Q}_{M_H}]^*_{,\R{g}}$ is slightly more
involved; it is obtained by differentiating $2 M_H^2$ with respect to
$\R{g}$ using \eqref{eq:FH}:
\begin{equation}
  \label{eq:11}
  \frac{\partial}{\partial \R{g}} (2M_H^2) 
  =
  -\frac{1}{\R{g}} \R{G}_{M_H} = -\frac{1}{\R{g}^*} [\R{G}_{M_H}]^* + {\cal O}(\delta g^2)
\end{equation}
as well as the right hand side (RHS) of Eq.~\eqref{eq:master_b},
\begin{eqnarray}
  \label{eq:22}
  \frac{\partial}{\partial \R{g}} (2M_H^2) &=& 
  \left( \frac{\beta}{2 \R{g}}\right)' \R{G}_{M_H} +  
  \left( \frac{\beta}{2 \R{g}}\right) \R{G}_{M_H}' +    \nonumber \\
  && ~~~~~~+ \gamma_m'  
  \R{Q}_{M_H} + (1+ \gamma_m^*)   \R{Q}_{M_H}'  + O(\delta \R{g}) \nonumber  \\
  &=&  \frac{\beta_1}{2 \R{g}^*} [\R{G}_{M_H}]^* +   
  \R{\gamma}_m^{(1)} [\R{Q}_{M_H}]^*  +
  (1+ \gamma_m^*) [\R{Q}_{M_H}]^*_{,\R{g}} + O(\delta \R{g})  \;,
\end{eqnarray}
where $'$ denotes differentiation with respect to $\R{g}$.  We have
dropped the term $\sim \beta \R{G}_{M_H}'$ from passing from the
first to the second line since it is of relative order ${\cal
  O}(\delta \R{g})$.  By equating Eqs.~\eqref{eq:11} and~\eqref{eq:22} we
may solve for $[\R{Q}_{M_H}]^*_{,\R{g}}$ and insert it into
\eqref{eq:dQ} and finally into \eqref{eq:indirect} to obtain:
 \begin{eqnarray}
   \label{eq:finito}
   \delta(2M_H^2) &=& \delta \R{g} b^{-\beta_1} 
   \left(  \frac{\beta_1}{2 \R{g}^*} [\R{G}_{M_H}]^* +
     \R{\gamma}_m^{(1)} [\R{Q}_{M_H}]^* \right)  + \nonumber   \\[0.1cm] 
   &+&  \delta \R{g} \Big(
   - 2 \frac{\R{\gamma}_m^{(1)}}{\beta_1}[ \R{Q}_{M_H}]^* f(b)  
   \underbrace{
     -\frac{1}{\R{g}^*}[\R{G}_{M_H}]^* b^{-\beta_1}  
     - b^{-\beta_1} \left( 
       \frac{\beta_1}{2 \R{g}^*} [\R{G}_{M_H}]^* + 
       \R{\gamma}_m^{(1)} [\R{Q}_{M_H}]^*\right) 
   }_{
     [\R{Q}_{M_H}]^*_{,\R{g}}
   } 
   \Big) \nonumber  \\[0.0cm]
   & = &  \delta \R{g} \left(  - \frac{1}{\R{g}^*} [\R{G}_{M_H}]^* b^{-\beta_1}   
     - 2 \frac{\R{\gamma}_m^{(1)}}{\beta_1} [\R{Q}_{M_H}]^* f(b) \right)  \;,
 \end{eqnarray}
 which equals Eq.~\eqref{eq:direct}, as expected.  We note that the second
 line in Eq.~\eqref{eq:finito} is equal to $(1+ \gamma_m^*) \delta
 \R{Q}_{M_H}$ at leading order.

 An interesting question is whether we can say something about the
 sign of the correction in Eq.~\eqref{eq:direct}.  That is to say we
 would like to know whether $l_{M_H}$ in $M_H^2 = k_{M_H} + l_{M_H}
 \delta \R{g}$ is positive or negative. We should add that $\delta g <
 0$ as can be inferred from Fig.~\ref{fig:one}. As we are interested
 in the long-distance dynamics of the theory that is defined at the UV
 gaussian fixed point, the coupling lies in the interval
 $\left[0,g^*\right]$. 
 
 As previously stated $\beta_1 >0$ in Eq.~\eqref{eq:ansatz} by virtue
 of no zero crossings of the $\beta$-function between the UV and IRFP.
 If the anomalous dimension increases monotonically from the UVFP
 $\gamma^*_{m,{\rm UV}} = 0$ to $\gamma^*_{m, { \rm IR}} = \gamma_m^*$
 then $\R{\gamma}_m^{(1)} > 0$, which is not compelling but to be
 expected.  Furthermore $[Q_{M_H}]^* > 0$ since $M_H^2 =
 (1+\gamma_m^*) [Q_{M_H}]^* + {\cal O}(\delta g)$ and $(1+\gamma_m^*)
 > 0$ as we have assumed $m$ to be a relevant direction and $f(b)
 < 0$ for $b>1$. Finally we see that everything depends on the sign of
 $G_{M_H}$ for which we cannot make a definite
 assertion.  It is well-known that naive positivity of operators,
 effective in quantum mechanics, is not necessarily maintained in
 quantum field theory. In the case at hand there is the additional
 complication that only the connected part of the matrix element is
 required.  That is to say even if the total matrix element were
 positive the connected part might still be negative. It seems
 worthwhile to point out that in QCD for $m \to 0$ and $\beta < 0$
 Eq.~\eqref{eq:master_b} implies that $G_{M_H} < 0$ indeed. Summa
 summarum we cannot say anything definite about the sign of $l_{M_H}$
 as the sign of $G_{M_H}$ seems uncertain.

\section{Conclusions}
We have explored the consequences of conformal scaling in a number of
interesting cases. 
One of our main findings is the scaling of the
radius of the $m$-hadrons as a function of the fermion mass. Our
results show that the typical size of the $m$-hadrons, defined from
the average charge density, is a linear function of the inverse mass
of the hadron \eqref{eq:r2H}. Characterising the size of $m$-hadrons is very
important in order to understand how to tame FSE in numerical studies,
and hence obtain reliable results from Monte Carlo simulations. It is
worthwhile to emphasise that the dependence of the mean charge radius
on the mass of the $m$-hadrons is radically different from the
logarithmic scaling obtained in chiral perturbation theory for the
Goldstone boson in a chirally broken theory \cite{GL85.III}.  
The difference between the two provides yet another way to asses 
the difference between a conformal and a confining phase.
 
 By exploiting selection rules for scaling dimensions and spin we propose 
 to use coordinate space correlation function, deformed by a mass term,
 to distinguish CFTs from SFTs  as well as confining theories. 
 
We investigated the scaling corrections to  correlation functions by 
linearizing the RGE in the variable $\delta g = g - g^*$ which is the distance of the initial  coupling from the, presumably unknown, fixed point value.  In essence this corresponds to 
the scaling corrections due to the IR-irrelevant coupling $g$.
The generic result is given in Eq.~\eqref{eq:schematic} 
and \eqref{eq:AB}. In subsection \ref{sec:discuss} we note in particular that scaling corrections can be expected to be large at the lower edge of the conformal window. 
This can be counteracted by going to smaller masses. 
We computed the scaling corrections to the hadron mass explicitly, once directly 
through its associated correlation function and second through the trace anomaly. 
The results are given in Eqs.~\eqref{eq:direct} \eqref{eq:indirect} and  their equivalence is made manifest in Eq.~\eqref{eq:finito}.  The latter was established 
by using the the Feynman-Hellmann relation for the mass and an analogous relation for the gauge coupling \eqref{eq:FH}. The derivation of the latter is given in a separate paper \cite{gluon-relations}.

\vspace{.5truein} {\bf Acknowledgements:} RZ acknowledges the support
of an advanced STFC fellowship. LDD and RZ are supported by an STFC
Consolidated Grant.  We are grateful to Julius Kuti, Jeff Fortin,
Martin L\"uscher and Tim Morris for discussion at various stages of
our investigations.

\appendix
\setcounter{equation}{0}
\renewcommand{\theequation}{A.\arabic{equation}}
\section{Charge and charge radius of pion form factor}
\label{app:charge}

In this appendix we shall give a derivation of the charge radius in terms of 
the form factor as stated in Eq.~\eqref{eq:pion-observables} as the derivation 
of the latter has become sparse in modern textbooks. We shall work in 
Minkowski-space in this section with metric signature $(+,-,-,-)$.
Starting from the zeroth component of \eqref{eq:FF}
\begin{equation}
\label{eq:again}
\matel{H(p_1)}{V_0(y)}{H(p_2)} = (E_{p_1}+ E_{p_2}) f^H_+(q^2) e^{i(p_1-p_2)\cdot y }      \;,
\end{equation}
where $E_p = \sqrt{\vec{p}^2+M_H^2}$ and $q \equiv p_1 - p_2$ as usual. 
Note for on-shell states the equality of the $3$-vectors $\vec{p_1} = \vec{p_2}$ then implies the vanishing of the $4$-vector $q = 0$.
We define the $D-1 = 3$-dimensional Fourier transform of the form factor
\begin{equation}
\label{eq:Fourier}
f^H(q^2) = \int \frac{d^3 x}{(2 \pi)^3} \hat f^H(\vec{x},q_0^2) e^{i \vec{x} \cdot \vec{q}} \;,
\end{equation}
for latter convenience. The scalar product with arrow vector denotes the $3$-dimensional scalar product.

\subsection{Charge}

The charge of the state $H$ is obtained by integrating the charge density over the 
space 
\begin{eqnarray}
\int d^3 x  \matel{H(p_1)}{V_0(x)}{H(p_1)}  &=& 2 E_p Q_H \big(  (2\pi)^3 \delta^{(3)}(0) \big) \;,  \nonumber \\[0.1cm] 
&\stackrel{\eqref{eq:again}}{=}& 2 E_p f_+^H(0)  
\big( \int_V d^3 x \big) \;,
\end{eqnarray}
remembering the normalisation $\vev{H(p_1)|H(p_2)} = 2 E_{p_1} (2 \pi)^3 \delta^{(3)}(\vec{p}_1-\vec{p})_2)$, setting $\vec{p_1} = \vec{p_2}$ on the first line and using the definition on 
the second line. To the more mathematical inclined reader this equation might look better
if $\vec{p}_1 = \vec{p}_2$ is not assumed before identifying $\int_V d^3 x =  (2 \pi)^3\delta^{(3)}(0)$. 
The latter identification  leads to the first result of this appendix:
\begin{equation}
\label{eq:1st}
\Rightarrow \quad f_H(0) = Q_H   \;,
\end{equation}
and suggests that 
\begin{equation}
\label{eq:use}
\hat f_H(\vec{x},0)/(2\pi)^3 = \rho(\vec{x})
\end{equation}
 is the charge density which we shall 
use below.

\subsection{Charge radius}
\label{app:charge}

We shall define the $3$-dimensional Laplace operator $\Delta$ 
acting on Fourier space as follows:
\begin{equation}
\label{eq:Laplace}
\Delta \circ {\cal F}(q) =  \sum_{a=1}^3 i \frac{d}{d q_a} i \frac{d}{d q_a}  {\cal F}(q)|_{q = 0} \;.
\end{equation}
We let it act on the form factor directly and through its Fourier transform:
\begin{eqnarray}
\label{eq:r2}
\Delta \circ f^H(q^2) &\stackrel{\eqref{eq:Laplace}}{=}& 
\left( 6 \frac{ d}{d q^2} f^H(q^2) + 4 \vec{q^2}  \frac{ d^2}{d (q^2)^2} f^H(q^2) \right) |_{q=0} = 
6 \frac{ d}{d q^2} f^H(q^2)|_{q=0}  \nonumber \\[0.1cm] 
&\stackrel{\eqref{eq:Fourier}}{=}&  
\int \frac{d^3 x}{(2 \pi)^3} \vec{x}^2 \hat f^H(\vec{x},q_0^2) e^{i \vec{x} \cdot \vec{q}} |_{q =0} = \int \frac{d^3 x}{(2 \pi)^3} \vec{x}^2 \hat f^H(\vec{x},0)  \;.
\end{eqnarray}
This leads, using \eqref{eq:use}, to the second result of this appendix:
\begin{equation}
\label{eq:2nd}
\Rightarrow \quad \vev{r^2_H} =   \int d^3 x \vec{x}^2 \rho_H(\vec{x})  \stackrel{\eqref{eq:r2}}{=}  6 \frac{ d}{d q^2} f^H(q^2)|_{q=0}  \;.
\end{equation}
Thus we have now justified the results quoted in Eq.~\eqref{eq:pion-observables} 
through \eqref{eq:1st} and \eqref{eq:2nd}.

\section{Finite size effects}
\label{app:FSE}

The aim of this appendix is to present extension of presentation in the main 
text to include finite size effects. 

\subsection{Generic two point function}
\label{app:genFSE}

Finite-size effects  to Eq.~\eqref{eq:Gtof}  can be easily incorporated. Writing explicitly the dependence of the correlators on the
size $L$ of the physical volume, the RG equation becomes: 
\begin{equation}
  \label{eq:RG-FSS}
  C(x,\hat m, \Lambda, L) = b^{-(\gamma^*_{O_1}+\gamma^*_{O_2})} C(x,
  b^{y^*_m} \hat m , \Lambda/b, L)  \;, \quad y^*_m = 1 + \gamma^*_m\, .
\end{equation}
The underlying assumption in the equation above is that the volume is
large enough, such that a blocking transformation does not change the
volume dependence. When all dimensionful quantities are rescaled by the
corresponding power of the reference mass $m_0$, Eq.~\eqref{eq:Gtof}
becomes:
\begin{equation}
  \label{eq:GtofFSS}
  C(x,\hat m,\Lambda,L) = \left(\hat x^2 \right)^{-\alpha}  \,
  (m_0)^{d_{O_1}+d_{O_2}+ d_{\varphi_a} + d_{\varphi_b} }  \, 
  F(\hat x^{y^*_m} \hat m, \hat \Lambda, \hat{x}/\hat{L})\, .
\end{equation}
In the thermodynamic limit, $\hat{L}\to\infty$
\begin{equation}
  \label{eq:Lexp}
  F(\hat x^{y^*_m} \hat m, \hat \Lambda , \hat{x}/\hat{L}) \to F (\hat
  x^{y^*_m} \hat m, \hat \Lambda) + \kappa \frac{\hat{x}}{\hat{L}} + \ldots \,,
\end{equation}
where $\kappa$ is a number.

\subsection{Charge radius}
\label{app:chargeFSE}

We  discuss the modifications of the scaling laws  
of the form factor \eqref{eq:FF}, relevant to the charge radius, 
due to finite-size effects. The form factor depends on the fermion mass, the
UV cut-off, and the physical size of the lattice:
\begin{equation}
  \label{eq:ffFSS}
  f(q^2) = f(q^2;\hat{m}, \Lambda, L)\, ,
\end{equation}
where  the hat indicates that dimensionful quantities have
been rescaled by the appropriate powers of the reference mass
$m_0$. Keeping $\Lambda$ unchanged, and performing the standard RG
analysis that we used above, yields:
\begin{equation}
  \label{eq:ffscal}
  f(q^2; \hat{m},\Lambda,L) =
  \tilde{f}(\frac{\hat{q}^2}{\hat{m}^{2/y_m}},\hat{L} \hat{m}^{1/y_m})\, .
\end{equation}
Expanding Eq.~(\ref{eq:ffFSS}) in powers of
$\hat{q}^2{\hat{m}^{-2/y_m}}$: 
\begin{equation}
  \label{eq:1FSSexpn}
  f(q^2; \hat{m},\Lambda,L) =
  \tilde{f}(0,\hat{L} \hat{m}^{1/y_m})+ \tilde{f}^\prime(0,\hat{L}
  \hat{m}^{1/y_m}) \frac{\hat{q}^2}{\hat{m}^{2/y_m}} + \ldots \, .
\end{equation}
This is the same expansion obtained in Eq.~(\ref{eq:taylor}), but now
the coefficients of the expansion depend on the physical size of the
lattice $L$. Denoting the $n$-th derivative of the form factor by
$\tilde{f}_{,n}$, and introducing the dimensionless finite-size
scaling variable $\ell=\hat{L} \hat{m}^{1/y_m}$, we obtain: 
\begin{equation}
  \label{eq:ffLexp}
  \tilde{f}_{,n}(0,\ell) = \frac{1}{\hat{L}^{y_m \eta_n}} \,
  \ell^{\eta_n y_m}  
  \left(   1 + \frac{\kappa}{\ell} +\ldots
  \right)
\end{equation}
with $\kappa$ a number and the dots denote the finite volume corrections. The dependence in Eq.~\eqref{eq:ffLexp}
reproduces the expected mass scaling discussed  before in the
large-volume limit.

\end{document}